\documentclass[reprint,showpacs,amssymb,amsmath,aps,pra]{revtex4-1}
\usepackage{epsfig}%
\newcommand{\half}{\mbox{$\frac{1}{2}$}}
\begin{document}
\title{Generalized entropic criterion for separability}
\author{R. Rossignoli$^1$, N. Canosa$^1$}
\affiliation{$^1$ Departamento de F\'{\i}sica, Universidad Nacional de La
Plata, C.C.67, La Plata (1900), Argentina}

\begin{abstract}
We discuss the entropic criterion for separability of compound quantum systems
for general non-additive entropic forms based on arbitrary concave functions
$f$. For any separable state, the generalized entropy of the whole system is
shown to be not smaller than that of the subsystems, for {\it any} choice of
$f$, providing thus a necessary criterion for separability. Nevertheless, the
criterion is not sufficient and examples of entangled states with the same
property are provided. This entails, in particular, that the conjecture about
the positivity of the conditional Tsallis entropy for all $q$, a more stringent
requirement than the positivity of the conditional von Neumann entropy, is
actually a necessary but not sufficient condition for separability in general.
The direct relation between the entropic criterion and the largest eigenvalues
of the full and reduced density operators of the system is also discussed.
\end{abstract}
\maketitle

The concept of quantum entanglement \cite{S.35} has aroused great interest in
recent years, due to its deep implications for quantum computation
\cite{Di.95}, quantum cryptography \cite{Ek.91} and quantum teleportation
\cite{Be.93}. The relation between entropy and quantum entanglement has also
attracted the attention from several authors
\cite{HHH.96,CA.97,HHH.99,BZ.99,AR.99,VB.99,AR.01,TLB.01,TPA.02,GG.01,CR.02}.
It is well known, for instance, that the von Neumann entropy of a compound
quantum system may be larger or smaller than that of a subsystem
\cite{LR.73,W.78}. However, if the system is in a separable (i.e., unentangled)
state, the von-Neumann entropy of the whole system is not smaller than that of
a subsystem \cite{HHH.96,CA.97}. Unfortunately, the converse is not true, i.e.,
the same may occur when the system is in an inseparable (i.e., entangled)
state, so that this entropy provides only a necessary test for separability.
The von Neumann based criterion is actually rather weak, being less stringent
than other equally simple necessary conditions \cite{HHH.96,P.96,HHH2.96}. As
discussed in \cite{HHH.99,GG.01,CR.02}, the von Neumann entropy is in fact not
a good entanglement indicator even in those cases where entanglement is fully
determined by the eigenvalues of the density operator $\rho$.

These facts suggest consideration of other information measures which could capture
more effectively the effects associated with the separability or inseparability
of a compound quantum system. In particular, it has been shown that
non-additive information measures like that of Tsallis \cite{T.88} do provide
more stringent conditions for separability \cite{AR.01,TLB.01}. Moreover, this
entropy depends on a parameter $q$ which can be optimized. In fact, for
$q\rightarrow\infty$, necessary and sufficient conditions for separability were
obtained with this entropy  for some important classes of states, like Werner
states for $n$ {\it qubits} and also $n$ {\it qudits} \cite{AR.01,TLB.01}. In
other situations \cite{TPA.02}, entanglement was detected however at {\it
finite} values of $q$, rather than in the $q\rightarrow\infty$ limit. Hence,
the questions arise of whether this entropy could provide a necessary and/or 
sufficient test in general and whether other information measures could lead to 
the same result. 

In this article we will examine more general entropic forms based on arbitrary
concave functions, which include as particular cases the von Neumann and
Tsallis entropies. We will show that any of these forms provide {\it necessary}
conditions for separability, which are not sufficient in general. It will also
become clear why the Tsallis form provides necessary and sufficient conditions
for Werner states in the $q\rightarrow\infty$ limit, and why it is not so in
other situations. Finally, other entropic forms providing similar results
are given.

Let us consider a quantum system described by a density operator $\rho$. We
will examine the general entropic forms \cite{CR.02}
 \begin{equation}
S_f(\rho)={\rm Tr}\,f(\rho)=\sum_i f(p_i)\,,\label{1}
 \end{equation}
where $f$ is a smooth {\it concave} function ($f'(p)$ decreasing for $p\in(0,1)$)
satisfying $f(0)=f(1)=0$, and $p_i$ , $i=1,\ldots,n$, are the eigenvalues of
$\rho$ ($\sum_ip_i=1$). We assume a {\it finite} dimension $n$. The von
Neumann entropy is recovered for 
 \begin{equation}
f(p)=-k p\ln p\,,\label{Sh}
 \end{equation}
with $k>0$, while the Tsallis entropy corresponds to \cite{T.88}
 \begin{equation}
f(p)=(p-p^q)/(q-1),\;\;\;q>0\label{Ts}
 \end{equation}
which approaches  $-p\ln p$ for $q\rightarrow 1$. The generalized entropies
(\ref{1}) satisfy most basic properties of the conventional entropy, except
those related with additivity. In particular, $S_f(\rho)\geq 0$, with
$S_f(\rho)=0$ iff the system is in a pure state ($\rho^2=\rho$), while its
maximum is attained for the fully mixed state $\rho=I/n$ \cite{RC.99}.
Concavity of $f$ ensures concavity of $S_f(\rho)$ \cite{W.78} ($S_f(\sum_j
q_j\rho_j)\geq \sum_jq_jS_f(\rho_j)$ for $0\leq q_j\leq 1$, $\sum_jq_j=1$). It
can be shown \cite{RC.99,CR.02} that if $[pf''(p)]'$ $\leq 0$ ($\geq 0$), then
$S_f$ is sub(super)-additive, i.e., $S_f(\rho_A\otimes\rho_B)-S_f(\rho_A)-
S_f(\rho_B)\leq 0$ ($\geq 0$). The condition $[pf''(p)]'=0$ determines in fact
Eq.\ (\ref{Sh}). The Tsallis entropy is, accordingly, sub(super)-additive for
$q>1$ ($q<1$).

A fundamental property of the forms (\ref{1}) which will be employed in this
work, and which justifies their use as information measures, is that if $\rho$
is {\it more mixed} than a density operator $\rho'$, then
 \begin{equation}
S_f(\rho)\geq S_f(\rho')\,,\label{11}
 \end{equation}
for {\it any} $f$ of the previous form \cite{W.78}. Labeling the eigenvalues of
$\rho$ and $\rho'$ in {\it decreasing} order, i.e. $p_1\geq p_2\geq\ldots \geq
p_n$, $\rho$ is said to be more mixed (or disordered) than $\rho'$ if
 \begin{equation}
{\cal S}_i=\sum_{j=1}^ip_j\leq{\cal S}'_i=\sum_{j=1}^ip'_j\,,\;\;\;
i=1,\ldots,n-1\,,\label{in}
 \end{equation}
i.e., if $p_1\leq p'_1$, $p_1+p_2\leq p'_1+p'_2$, etc (for $i=n$, ${\cal
S}_n={\cal S}_n'=1$). Mathematically, this states that the set of probabilities
$(p_1,\ldots,p_n)$ is {\it majorized} by $(p'_1,\ldots,p'_n)$. Eq.\ (\ref{11})
can be immediately derived writing $p_i={\cal S}_i-{\cal S}_{i-1}$ in
(\ref{1}), with ${\cal S}_0=0$. $S_f(\rho)$ is then a {\it decreasing} function
of ${\cal S}_i$ for  $1\leq i\leq n-1$, as $\partial S_f/\partial{\cal
S}_i=f'(p_i)-f'(p_{i+1})\leq 0$ if $p_i\geq p_{i+1}$ and $f$ is concave (Eq.\
(\ref{11}) follows then from the mean value theorem; note that the allowed
values of ${\cal S}_i$ form a convex set defined by ${\cal S}_i\leq {\cal
S}_{i+1}$, ${\cal S}_{i}\geq ({\cal S}_{i-1}+{\cal S}_{i+1})/2$, with ${\cal
S}_0=0$, ${\cal S}_n=1$).

Moreover, it can be shown \cite{W.78} that $\rho$ is more mixed than $\rho'$
{\it if and only if} ${\rm Tr}\,f(\rho)\geq {\rm Tr}\,f(\rho')$ for {\it any}
concave $f$, i.e., iff Eq.\ (\ref{11}) holds $\forall$ $f$ of the previous form
(the conditions $f(0)=f(1)=0$ fix just an arbitrary linear term $ap+b$ that can
be added to $f$ without affecting concavity or Eq.\ (\ref{11})). If the
dimensions of $\rho$ and $\rho'$ differ, we may apply the same definition of
more mixed by adding zero eigenvalues to the density with the smallest
dimension, which leaves $S_f$ unchanged.

Let us consider now a system composed of two subsystems $A$ and $B$. The
quantity
 \begin{equation}
S_{f}^A(\rho)\equiv S_f(\rho)-S_f(\rho_A)=
{\rm Tr}\,f(\rho)-{\rm Tr}_A\,f(\rho_A)\,,\label{2}
 \end{equation}
where $\rho_A={\rm Tr}_B\,\rho$ is the reduced density matrix of system $A$ and
${\rm Tr}={\rm Tr}_A{\rm Tr}_B$, plays the role of a {\it conditional} entropy.
In the von Neumann case, Eq.\ (\ref{2}) becomes the usual conditional entropy
\cite{W.78},
 \[S_f^A(\rho)=S(B|A)=-{\rm Tr}\,\rho[\ln\rho-\ln\rho_A\otimes I_B]\,,\]
whereas in the Tsallis case, it is proportional to the $q$-conditional entropy
defined in \cite{AR.01,TLB.01}, $S_q(B|A)=S_f^A(\rho)/{\rm Tr}\,\rho_A^q$.

For a {\it discrete classical system} described by a joint probability
distribution $p_{ij}$, Eq.\ (\ref{2}) is always {\it non-negative}, i.e.,
 \begin{equation}
\sum_{i,j}f(p_{ij})-\sum_if(p_i)\geq 0,\;\;\;p_i=\sum_{j}p_{ij}\,,
 \end{equation}
since for {\it any} concave $f$ satisfying $f(0)=0$, we have  $f(p+q)\leq
f(p)+f(q)$ if $p\geq 0$, $q\geq 0$ (it may be also seen that the set of
probabilities $\{p_{ij}\}$ is {\it more mixed} than $\{p_i\}$). This implies
that $S^{A}_f(\rho)\geq 0$ for any uncorrelated density
$\rho=\rho_A\otimes\rho_B$ (i.e. $p_{ij}=p^A_ip^B_j$) as well as for any
density diagonal in a basis of product states
($\rho=\sum_{i,j}p_{ij}|i_Aj_B\rangle\langle i_Aj_B|$). Nevertheless, in the
general quantum case, $S_f^{A}(\rho)$ may of course be negative. In particular,
for a pure state $\rho=|\Psi\rangle\langle\Psi|$, $S_f(\rho)=0$ and the
positive eigenvalues of $\rho_A$ and $\rho_B$ are identical \cite{W.78}, whence
 \begin{equation}
S_f^{A}(\rho)=-S_f(\rho_A)=-S_f(\rho_B)\leq 0\,.
 \end{equation}
For $f(p)=-p\,{\rm log}_2\,p$,  this is just the usual definition of the
{\it entanglement} of a pure state $|\Psi\rangle$ \cite{BB.96,W.98}.

Negative values of $S^{\rm A}_f(\rho)$ are then indicative of distinctive
quantum correlations. In particular, for the case (\ref{Ts}) it has been
conjectured \cite{AR.01,TLB.01,TPA.02} that the sign of the difference
(\ref{2}) may provide a criterion for determining the {\it separability} of
$\rho$ \cite{TPA.02}. Let us recall that a mixed state $\rho$ is {\it
separable} (or clasically correlated) iff it can be written as a convex
combination of uncorrelated densities \cite{W.89},
 \begin{equation}
\rho=\sum_\alpha \omega_\alpha \rho^{\alpha}_A\otimes\rho^{\alpha}_B\,,
\;\;\;\;\;0\leq\omega_\alpha\leq 1,
 \end{equation}
with $\sum_\alpha\omega_\alpha=1$. Otherwise it is called {\it entangled} or
{\it inseparable}. For the Tsallis case, it has been shown \cite{AR.01,TLB.01}
that the criterion $S_f^A(\rho)\geq 0$ leads, for $q\rightarrow\infty$, to the
necessary and sufficient condition for separability for some important classes
of states, like Werner states. Nevertheless, we will show here that this does
not hold in general. In particular, for an entangled state $S_f^A(\rho)$ and
$S_f^B(\rho)$ may in fact be both positive {\it for any} concave $f$ (including
the $q\rightarrow\infty$ limit in the Tsallis case), indicating that
entanglement cannot be always detected by such entropic criteria (or, in
general, by information based on the eigenvalues of $\rho$ and $\rho_{A,B}$
alone). This may occur already for a two qubit system, where the Peres
necessary criterion for separability \cite{P.96} is known to be sufficient
\cite{HHH2.96}, so that the entropic criterion is here weaker than the Peres
criterion.

Let us first show that Eq.\ (\ref{2}) is indeed positive for {\it any}
separable $\rho$. A fundamental theorem demonstrated  in \cite{N.01} states
that if $\rho$ is {\it separable}, then $\rho$ {\it is more mixed than $\rho_A$
and $\rho_B$} (disorder criterion for separability). Hence, Eq.\ (\ref{11})
implies that if $\rho$ is separable, then
 \begin{equation}
S_f^A(\rho)\geq 0\,,\label{Sfe}
 \end{equation}
and similarly, $S_f^B(\rho)\geq 0$, {\it for any concave} $f$ (satisfying
$f(0)=0$). This is in fact an equivalent entropic formulation of the disorder
criterion. For a separable state, Eq.\ (\ref{Sfe}) will therefore hold $\forall
q>0$ in the case (\ref{Ts}), implying ${\rm Tr}\,\rho^q-{\rm
Tr}_A\,\rho_A^q\leq 0$ ($\geq 0$) if $q>1$ ($0<q<1$). Note that this entails
$S_\alpha(\rho)\geq S_\alpha(\rho_A)$ $\forall$ $\alpha>0$, where
$S_\alpha(\rho)=\frac{1}{1-\alpha}\ln {\rm Tr}\, \rho^\alpha$ is the {\it
R\'enyi} entropy \cite{R.70,HHH.96} (which is additive but not of the form
(\ref{1}), and approaches the von Neumann entropy for $\alpha\rightarrow 1$).
The disorder criterion is, however, {\it not} sufficient \cite{N.01}, so that
Eq.\ (\ref{Sfe}) provides in general only a necessary test for separability, as
will be explicitly seen below.

For a system of two qubits, Eq.\ (\ref{Sfe}) is actually an immediate
consequence of the more obvious fact that for {\it any} separable state,
 \begin{equation}
p_1\leq p_1^A\,,\label{pa}
 \end{equation}
where $p_1$ ($p_1^A$) denotes the {\it largest} eigenvalue of $\rho$
($\rho_A$). This is so because the difference
 \begin{equation}
\rho_d=\rho_A\otimes I_B-\rho=
\sum_\alpha\omega_\alpha\rho_A^\alpha\otimes(I_B-\rho_B^\alpha)\,,
 \end{equation}
is a {\it non-negative} operator if all $\omega_\alpha\geq 0$ \cite{CAG.99}.
Hence, denoting with $|i\rangle$ any eigenstate of $\rho$, we have
 \begin{equation}
0\leq \langle i|\rho_d|i\rangle=\langle i|\rho_A\otimes I_B|i\rangle-p_i
\leq p^A_{1}-p_i\,,\label{des}
 \end{equation}
since $\langle i|\rho_A\otimes I_B|i\rangle\leq \langle 1_Aj_B|\rho_A\otimes
I_B|1_Aj_B\rangle=p_1^A$, where $\rho_A|1_A\rangle=p_1^A|1_A\rangle$ and
$|j_B\rangle$ is any state of $B$. For a two qubit system, (\ref{pa}) already
implies that $\rho$ is {\it more mixed} than $\rho_A$: $\sum_{j=1}^i p_j\leq
p_1^A+p_2^A=1$ for $i=2,3,4$.

There are two important remarks to make here. First, if $p_1>p_1^A$, the state
is certainly entangled, but $\rho_A$ is not necessarily more mixed than $\rho$,
entailing that $S_f^{A}(\rho)$ is not necessarily negative for any $f$.
Nevertheless, in the Tsallis case, as well as for any set of entropic functions
 \begin{equation}
f(p)=k[p-g_q(p)]\,,\label{gq}
 \end{equation}
where $k>0$ and $g_q(p)$ is a convex {\it increasing} function satisfying
$g_q(0)=0$, $g_q(1)=1$ and
 \begin{equation}
\lim_{q\rightarrow\infty}g_q(p')/g_q(p)=0\;{\rm if}\;p'<p\,,\label{lim}
 \end{equation}
$S_f(\rho)$ will be a {\it decreasing} function of the largest eigenvalue $p_1$
for sufficiently large $q$ and finite dimension
($S_f(\rho)\approx k(1-d_1g_q(p_1))$ in this limit, with $d_1$ the multiplicity
of $p_1$). Hence, if $p_1>p_1^A$, $S_f^{A}(\rho)$ {\it will become negative}
for sufficiently large $q$, and the entropic criterion will be able to detect
entanglement. In other words, for $q\rightarrow\infty$, $S_f^{A}(\rho)<0$ iff
$p_1>p_1^A$, which is a {\it sufficient} condition for inseparability. Note
that Eq.\ (\ref{Ts}) is of the form (\ref{gq}) for $q>1$ and satisfies
(\ref{lim}). Another example is \cite{CR.02}
 \begin{equation}
f(p)=[p-\frac{e^{qp}-1}{e^{q}-1}]/q\,,\label{ex}
 \end{equation}
which is concave $\forall$ $q$, approaches $\half p(1-p)$ for $q\rightarrow 0$
($q=2$ case in (\ref{Ts})) and is of the form (\ref{gq}) for $q>0$.

Nonetheless, and this is the second important remark, there are entangled
states for which $p_1\leq p_1^{A}$ and $p_1^B$, i.e., for which the greatest
eigenvalue of $\rho$ remains smaller than that of $\rho_A$ and $\rho_B$. This
may occur already for a system of two qubits, in which case $\rho$ will remain
more mixed than $\rho_A$ and $\rho_B$, and $S_f^{A}(\rho)$, $S_f^B(\rho)$ will
both be non-negative for {\it any} concave $f$. This type of entanglement will
therefore not be detected by the previous entropic criterion.

\begin{figure}[t]
\vspace*{-4cm}

\hspace*{-2cm}\scalebox{0.7}{\includegraphics{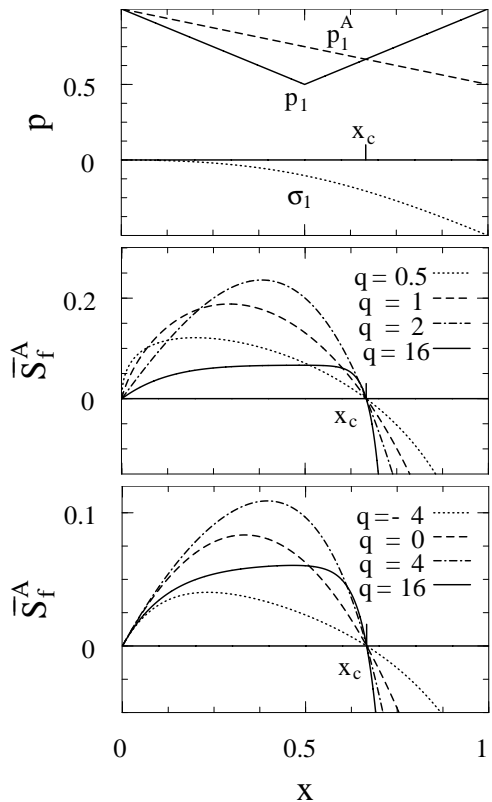}}\vspace*{-5.75cm}

\caption{Top: The largest eigenvalue $p_1$ of $\rho$ and $p_1^A$ of $\rho_A$,
for the density (\protect\ref{s1}), as a function of the parameter $x$. The
dotted line depicts the lowest eigenvalue $\sigma_1$ of the partial transpose
of $\rho$. Center: The normalized entropic difference (\protect\ref{Sbar}) for
the Tsallis case (\protect\ref{Ts}), at the indicated values of $q$. The curve
for $q=1$ corresponds to the von Neumann entropy, in which case
$\bar{S}_f^A=S(B|A)$. Bottom: The same quantity for the entropic function
(\protect\ref{ex}). The curve for $q=0$ depicts the limit
$\bar{S}_f^A=\half{\rm Tr}[\rho^2_A-\rho^2]$. The point where $p_1=p_1^A$ is
indicated by $x_c$. Both $x$ and the quantities plotted are
dimensionless.}\label{f1}
\end{figure}

An example is the state considered in \cite{P.96},
 \begin{equation}
\rho=x|\Psi_0\rangle\langle\Psi_0|+(1-x)
|\!\!\uparrow\uparrow\rangle\langle\uparrow\uparrow\!\!|
\,,\;\;\;\;0\leq x\leq 1\,,\label{s1}
 \end{equation}
where $|\Psi_0\rangle=(|\!\!\uparrow\downarrow\rangle\!
-\!|\!\!\downarrow\uparrow\rangle)/\sqrt{2}$ is the singlet (a maximally
entangled state) and $|\!\!\uparrow\uparrow\rangle$ a maximally polarized
separable state. As shown in \cite{P.96}, Peres criterion determines that this
state is {\it entangled} $\forall$ $x>0$: the partial transpose of $\rho$
(defined as the transposition with respect to the indexes of system $A$), which
is still a density operator if $\rho$ is separable, has always a negative
eigenvalue for $x>0$, namely
$\sigma_1=\half(1\!-\!x\!-\!\sqrt{1\!-\!2x(1\!-\!x)})\;\;$
($\sigma_1=-x^2/4+O(x^3)$ for $x\rightarrow 0$).

However, as the eigenvalues of $\rho$ are $(x,1-x,0,0)$, and those of $\rho_A$
and $\rho_B$ are $(1-x/2,x/2)$, the greatest eigenvalue of $\rho$ ($p_1=x$ for
$x>\half$) is greater than that of $\rho_A$ ($p^A_1=1-x/2$) only for
$x>x_c=2/3$ [Fig.\ \ref{f1}]. Hence, for $0<x<2/3$, entanglement {\it will not
be detected by} $S_f^{A,B}(\rho)$, {\it for any} $f$. This can also be directly
seen from the explicit expression
 \[S_f^{A}(x)=f(x)+f(1-x)-[f(x/2)+f(1-x/2)]\,.\]
Since for a two state system, the entropy $f(p)+f(1-p)$ is a {\it decreasing}
function of the largest eigenvalue ($f'(p)-f'(1-p)<0$ for $p>1/2$ and $f$
concave), in this case $S_f^{A}(\rho)<0$ iff $p_1>p_1^A$, i.e.,  $S_f^{A}(x)<0$
iff $x>2/3$, for {\it any} $f$. The sign of $S_f^{A}(x)$ is {\it independent}
of the choice of entropic function $f$ in this example, i.e. independent of $q$
in the Tsallis case or in Eq.\ (\ref{ex}), as shown in Fig.\ \ref{f1}. For
normalization purposes, we have plotted the quantity
 \begin{equation}
\bar{S}_f^A(\rho)=S_f^A(\rho)/{\rm Tr}\,g_q(\rho_A)\,,\label{Sbar}
 \end{equation}
where $g_q(p)=p^q$ in the Tsallis case (\ref{Ts}) (so that
$\bar{S}_f^A(\rho)=S_q(B|A)$) and $g_q(p)=(e^{qp}\!-\!1)/(e^{q}\!-\!1)$ for
Eq.\ (\ref{ex}).

This situation is actually not very special. Consider for instance the more
general state
 \begin{equation}
\rho=x|\Psi_0\rangle\langle\Psi_0|+(1-x)|uv\rangle\langle uv|\label{s1g}\,,
 \end{equation}
where $|uv\rangle=|u\rangle_A|v\rangle_B$ is an arbitrary separable pure state
of the two qubits. This state is again {\it entangled} $\forall$ $x>0$, since
the partial transpose of $\rho$ has a negative eigenvalue
 \[\sigma_1=\half(1\!-\!x\!-\!\sqrt{1\!-\!2x(1\!-\!x)r}),\;\;\;\
 r=|\langle u|v\rangle|^2\,,\]
with $\sigma_1=-x(1\!-\!r)/2+O(x^2)$ for $x\rightarrow 0$.
On the other hand, the eigenvalues of $\rho$ are
 \[(\half(1+z),\half(1-z),0,0),\;\;z=\sqrt{1\!-\!2x(1\!-\!x)(1\!+\!r)}\,,\]
while those of $\rho_{A}$, $\rho_{B}$ are again $(1-x/2,x/2)$. Hence,
$p_1=(1+z)/2$, $p_1^A=(1-x/2)$, and $p_1>p_1^A$ only for
 \[ x>x_c=2r/(1+2r)\,. \]
Thus, $S_f^{A}(\rho)<0$ iff $x_c<x<1$, for {\it any concave} $f$. Again, the
entropic criterion fails to detect entanglement for $0<x<x_c$. For $r=1$, we
recover the results of the previous example, whereas for $r=0$, i.e.
$|uv\rangle=|\!\!\uparrow\downarrow\rangle$, $\sigma_1=-x/2$ and $x_c=0$, so
that $S_f^{A}(\rho)<0$ $\forall$ $x>0$. This is the only case where the
entropic criterion predicts the full interval of inseparability.

\begin{figure}[t]
\vspace*{-4cm}

\hspace*{-2cm}\scalebox{0.7}{\includegraphics{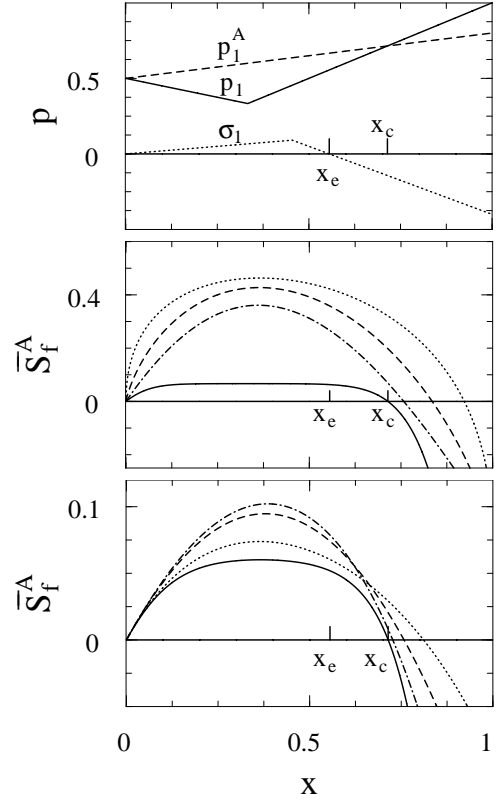}}\vspace*{-5.75cm}

\caption{Same details as Fig.\ 1 for the density (20) with $|a|^2=4/5$. 
The values of $q$ for the different lines in the center and
bottom panels are the same as those of Fig.\ 1.}
\label{f2}\end{figure}

Let us still consider the example of refs.\ \cite{G.96,P.96},
 \begin{equation}
\rho=x|\Psi\rangle\langle\Psi|+(1-x)(|\!\!\uparrow\uparrow\rangle\langle
\uparrow\uparrow\!\!|+|\!\!\downarrow\downarrow\rangle
\langle\downarrow\downarrow\!\!|)/2\,,\label{s3}
 \end{equation}
with $|\Psi\rangle=a|\!\!\uparrow\downarrow\rangle+b|\!\!\downarrow\uparrow
\rangle$, $|a|^2+|b|^2=1$. As shown in \cite{P.96}, this state is entangled
just for
 \[x>x_e=(1+2|ab|)^{-1}\,,\]
since the lowest eigenvalue of the partial transpose is
$\sigma_1=(1\!-\!x(1\!+\!2|ab|))/2$ for
$x>[2(1\!+\!|ab|)\!-\!||a|^2\!-\!|b|^2|]^{-1}$. However, the eigenvalues of
$\rho$ are $(x,(1\!-\!x)/2,(1\!-\!x)/2,0)$ while those of $\rho_{A}$, $\rho_B$
are $(1\!\pm\!x(|b|^2\!-\!|a|^2))/2$. The largest eigenvalue of $\rho$ ($p_1=x$
for $x>1/3$) is greater than that of $\rho_{A}$
($p_1^A=(1+x||a|^2\!-\!|b|^2|)/2)$ only for
 \[x>x_c=(2\!-\!||a|^2-|b|^2|)^{-1}\,.\]
But $x_c\geq x_e$, with $x_c=x_e$ just for $|a|=|b|$ or in the trivial
separable cases $b=0$ or $a=0$. Hence, if $|a|\neq |b|$ and $ab\neq 0$,
$S_f^{A}(\rho)$ will not detect entanglement for $x_e<x<x_c$. Note also that
for $x>x_c$, $S_f^A(\rho)$ is in this case not necessarily negative for any
$f$, but will become negative for sufficiently large $q$ in the Tsallis case or
in Eqs.\ (\ref{gq}) or (\ref{ex}), as shown in Fig.\ \ref{f2}. The value of $x$
where $S_f^A(\rho)=0$ converges actually exponentially fast to $x_c$ for
$q\rightarrow\infty$ in (\ref{Ts}) or (\ref{ex}). This will occur whenever the
degeneracies of $p_1$ and $p_1^A$ coincide.

The entropic criterion will provide, however, necessary and sufficient
conditions for separability for {\it any} density $\rho$ {\it diagonal in the
Bell basis} \cite{HHH.99}, i.e. the basis of maximally entangled states
$|\Psi_0\rangle$, $|\Psi_1\rangle=(|\!\!\uparrow\downarrow\rangle
+|\!\!\downarrow\uparrow\rangle)/\sqrt{2}$,
$|\Psi_{2,3}\rangle=(|\!\!\uparrow\uparrow\rangle
\pm|\!\!\downarrow\downarrow\rangle)/\sqrt{2}$. In such a case,
 \begin{equation}
\rho=\sum_{i=0}^3 q_i|\Psi_i\rangle\langle\Psi_i|\,,\label{rb}
 \end{equation}
is known to be entangled iff $p_1>1/2$ \cite{HHH.96}, where $p_1={\rm
Max}[\{q_i\}]$ is the largest eigenvalue of $\rho$. This may be obtained
directly with Peres criterion, as the partial transpose of $\rho$ has
eigenvalues $\half-q_i$. Now, for any pure Bell state
$|\Psi_i\rangle\langle\Psi_i|$, the reduced density matrices are {\it fully
mixed}, with eigenvalues $(\half,\half)$, so that the same will occur for any
state of the form (\ref{rb}). The condition $p_1\leq p_1^{A}$ then becomes 
equivalent, for {\it any} state (\ref{rb}), to $p_1\leq 1/2$, i.e., to the
necessary and sufficient condition for separability. The entropic criterion
will therefore always lead to this condition for $q\rightarrow\infty$ in
(\ref{gq}).

\begin{figure}[t]
\vspace*{-4cm}

\hspace*{-2cm}\scalebox{0.7}{\includegraphics{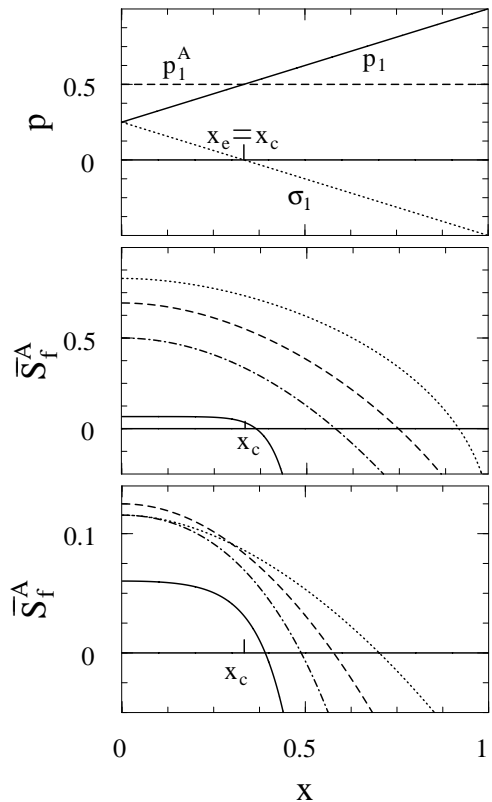}}\vspace*{-5.75cm}

\caption{Same details as Fig.\ 1 for the density (22).  
The values of $q$ for the different lines in the center and
bottom panels are the same as those of Fig.\ 1.}
\label{f3}\end{figure}

This explains why the entropic criterion for $q\rightarrow\infty$ yields the
necessary and sufficient condition for separability for Werner-Popescu states
\cite{W.89,Po.94},
 \begin{equation}
\rho=x|\Psi_0\rangle\langle\Psi_0|+(1-x)I/4\,,\label{s4}
 \end{equation}
where $I=\sum_{i=0}^3|\Psi_i\rangle\langle\Psi_i|=I_A\otimes I_B$ is the
identity. The eigenvalues of $\rho$ are $p_1=(1+3x)/4$ and $(1-x)/4$ (threefold
degenerate), and the equation $p_1\leq \half$ yields $x\leq\frac{1}{3}$, the
necessary and sufficient condition \cite{P.96,B.96}. Accordingly, for
$x>x_c=\frac{1}{3}$, $S_f^A(\rho)$ will become negative for sufficiently large
$q$. The root $x_r$ where $S_f^A(\rho)=0$ will approach $x_c$ for
$q\rightarrow\infty$, as seen in Fig.\ \ref{f3}, although the convergence is in
this case less rapid due to the different degeneracies of $p_1$ and $p_1^A$.
For large $q$,
 \begin{equation}
x_r\approx \frac{1}{3}+\frac{2\gamma\ln 2}{3q}\,,\label{xr}
 \end{equation}
where $\gamma=1$ for the Tsallis case and $\gamma=2$ for Eq.\ (\ref{ex}).

Note that the state (\ref{s3}) also becomes of the form (\ref{rb}) for $a=\pm
b$ (where the entropic criterion works), as in this case
$|\Psi\rangle=|\Psi_0\rangle$ or $|\Psi_1\rangle$ (the remaining term in
(\ref{s3}) is proportional to $\sum_{i=2,3}|\Psi_i\rangle\langle\Psi_i|$).

Similar considerations hold for Werner-like states for $n$ qubits
\cite{DCT.99},
 \begin{equation}
\rho=x|\Psi\rangle\langle\Psi|+(1-x)I/d^n\,,\label{s5}
 \end{equation}
where $d=2$, $|\Psi\rangle=(|\!\!\uparrow\uparrow\ldots\uparrow\rangle+
|\!\!\downarrow\downarrow\ldots\downarrow\rangle)/\sqrt{2}$ is a maximally
entangled state (a GHZ state \cite{GHZ.90}) and $I$ the identity. The
eigenvalues of (\ref{s5}) are $p_1=x+(1\!-\!x)/d^n$ and $(1\!-\!x)/d^n$
[$(d^n\!-\!1)$-fold degenerate]. Now, for a subsystem $A_m$ with $m$ qubits
($1\leq m \leq n-1$), the reduced density matrix $\rho_{m}$ can be easily shown
to have eigenvalues $p_1^{m}=x/d+(1\!-\!x)/d^m$ ($d$-fold degenerate) and
$(1\!-\!x)/d^m$ [($d^m-d$)-fold degenerate]. The necessary condition for
separability between the $m$ and $n-m$ subsystem, $p_1\leq p_1^{m}$, leads to
 \begin{equation}
x\leq x_c^m\equiv[1+\frac{d^{n-1}(d-1)}{d^{n-m}-1}]^{-1}\,,\label{xm}
 \end{equation}
which is a decreasing function of $m$. The most stringent condition is then
obtained for $m=n-1$, i.e., $x\leq(1+d^{n-1})^{-1}$, which, according to
refs.\ \cite{DCT.99,PR.00}, is just the {\it necessary and sufficient}
condition for full separability. The entropic criterion $S_f^{A_m}(\rho)\geq 0$
will then lead to Eq.\ (\ref{xm}) for $q\rightarrow\infty$ (as shown in
\cite{AR.01} for the Tsallis case). If $d$ is an arbitrary integer ($\geq 2$),
the previous discussion and expressions are actually also valid for $n$
{\it qudits} ($n$ $d$-dimensional systems), when $|\Psi\rangle$ is the fully
entangled state $\sum_{k=0}^{d-1}|k\rangle_1\ldots|k\rangle_n/\sqrt{d}$
\cite{PR.00}.

It should be stressed that for bipartite systems with subsystem dimension
$d>2$, the first violation of the majorization relation between $\rho$ and
$\rho_A$ in an entangled state may also occur for $i>1$ in Eq.\ (\ref{in}). For
instance, let us briefly discuss the example given in \cite{TPA.02}, dealing
with a system of two identical harmonic oscillators. It was shown that for
certain densities, $S_f^A(\rho)$ becomes negative just in a {\it finite}
interval of $q$ values in the Tsallis case, remaining positive for arbitrary
large $q$. This indicates that $\rho$ is not more mixed than $\rho_A$, and
hence entangled, {\it but still has} $p_1<p_1^A$, which ensures that
$S_f^A(\rho)$ remains positive for $q\rightarrow\infty$. The first violation of
the inequalities (\ref{in}) is therefore taking place for $i>1$ (we have
verified that this occurred for $i=2$). Nevertheless, it should be remarked
that in such situations, if ${\cal S}_i$ is only slightly larger than ${\cal
S}_i^A$ and $i>1$, $S_f^A(\rho)$ may remain positive for all $q>0$ in the case
(\ref{Ts}), being then unable to detect entanglement. The same happens with the
entropy (\ref{ex}).

In summary, we have shown that the generalized entropic criterion
$S_f^A(\rho)=S_f(\rho)-S_f(\rho_A)\geq 0$ constitutes, for {\it any} concave
entropic function $f$, a {\it necessary} condition for separability. For
$q\rightarrow\infty$ in Eq.\ (\ref{Ts}), or in general Eq.\ (\ref{gq}), it
becomes equivalent to the condition (\ref{pa}) between the largest eigenvalues
of $\rho$ and $\rho_A$. Nonetheless, the entropic criterion is not a sufficient
one in general. We have provided examples of entangled densities of two qubits
where $p_1<p_1^A$, in which case $\rho$ remains {\it more mixed} than $\rho_A$,
implying $S^A_f(\rho)\geq 0$ for {\it any} choice of entropic function $f$.
However, the condition $p_1\leq p_1^A$ becomes sufficient in some important
cases, which include {\it any} density diagonal in the Bell basis in a two
qubit system, and also Werner-like states in $n$ qubit (or qudit) systems. In
these cases the inequality $S_f^A(\rho)\geq 0$ will lead, for
$q\rightarrow\infty$ in Eq.\ (\ref{Ts}) or (\ref{gq}), to the necessary and
sufficient condition for separability.

The condition $S^A_f(\rho)\geq 0$ for {\it any} concave entropic function $f$
is equivalent to the requirement that $\rho$ be {\it more mixed} than $\rho_A$,
a general {\it necessary} condition for separability \cite{N.01}. Let us remark
that this requirement is {\it stronger} than the condition $S_f^A(\rho)\geq 0$
$\forall$ $q>0$ in (\ref{Ts}) (or $\forall q$ in (\ref{ex})). Other families of
concave entropic functions are required in general to detect that $\rho$ is not
more mixed than $\rho_A$ when the first violation of Eqs.\ (\ref{in}) occurs
for $i>1$, although in many cases this can also be seen with the entropies
(\ref{Ts}) or (\ref{ex}). In such situations $S_f^A(\rho)$ will remain positive
for $q\rightarrow\infty$ but may become negative at finite values of $q$.

RR and NC acknowledge support from CIC and CONICET, respectively, of Argentina.


\end{document}